\documentclass{article}

\usepackage{arxiv}

\usepackage[utf8]{inputenc} 
\usepackage[T1]{fontenc}    
\usepackage{url}            
\usepackage{booktabs}       
\usepackage{amsfonts}       
\usepackage{nicefrac}       
\usepackage{microtype}      
\usepackage{lipsum}

\usepackage[authoryear,longnamesfirst]{natbib}
\usepackage{float}
\usepackage{booktabs}

\usepackage{fancyhdr}       
\usepackage{multirow}

\usepackage{graphicx}       
\graphicspath{{media/}}     
\usepackage[dvipsnames]{xcolor}

\definecolor{color1}{rgb}{0.576,0.769,0.49}

\title{A Framework for Collaborating a Large Language Model Tool in Brainstorming for Triggering Creative Thoughts
}

\author{
  Hung-Fu Chang \\
  R.B. Annis School of Engineering \\
  University of Indianapolis \\
  Indianapolis\\
  \texttt{hchang@uindy.edu} \\
   \And
  Tong Li \\
  School of Journalism and Strategic Communication \\
  Ball State University \\
  Muncie\\
  \texttt{tong.li@bsu.edu} \\
}

\begin{document}
\maketitle

\begin{abstract}
Creativity involves not only generating new ideas from scratch but also redefining existing concepts and synthesizing previous insights. Among various techniques developed to foster creative thinking, brainstorming is widely used. With recent advancements in Large Language Models (LLMs), tools like ChatGPT have significantly impacted various fields by using prompts to facilitate complex tasks. While current research primarily focuses on generating accurate responses, there is a need to explore how prompt engineering can enhance creativity, particularly in brainstorming. Therefore, this study addresses this gap by proposing a framework called GPS, which employs goals, prompts, and strategies to guide designers to systematically work with an LLM tool for improving the creativity of ideas generated during brainstorming. Additionally, we adapted the Torrance Tests of Creative Thinking (TTCT) for measuring the creativity of the ideas generated by AI. Our framework, tested through a design example and a case study, demonstrates its effectiveness in stimulating creativity and its seamless LLM tool integration into design practices. The results indicate that our framework can benefit brainstorming sessions with LLM tools, enhancing both the creativity and usefulness of generated ideas.
\end{abstract}


\section{Introduction}
Creativity isn't just about creating new ideas from scratch; it also involves redefining existing thoughts and synthesizing previous ideas (Gafour et al., 2020). It is a process that includes becoming sensitive to problems, deficiencies, information shortages, missing elements, identifying challenges, searching for solutions, making estimates and hypotheses, modifying these hypotheses in response to the identified shortcomings, and trying one of these solutions (Torrance, 1974). These elements of the process can be formulated into techniques that encourage creative idea generation or creative thinking. Many techniques have been recommended by previous studies (Herrmann et al. 2018). Research has shown that using those techniques results in more innovative ideas than not using any technique at all in practice (Chulvi et al., 2012; Dumas et al., 2015).

One of the frequently used techniques is Brainstorming (Diehl \& Stroebe, 1991). It is an unstructured technique used in a group setting to generate creative ideas for a specific problem (Rossiter et al. 1994; Besant 2016). Brainstorming stimulates creativity and out-of-the-box thinking, which can lead to the exploration of unconventional solutions and the discovery of novel approaches to tackle a complex problem. It is also a versatile tool utilized in various fields, such as business and education. However, although barnstorming is unstructured in nature, it is important to have clear principles or tools to improve the group’s working efficiency and increase the creativity and quality of the idea generated (Rossiter et al. 1994).

Recent advances in Artificial Intelligence, particularly in the field of Natural Language Processing, have significantly impacted our lives through the development of Large Language Models (LLMs). Tools like ChatGPT have demonstrated their ability to assist users with a wide range of complex tasks. Their ability extends beyond information extraction (Shah et al. 2024) to data analysis, media creation, and reasoning (Chew et al. 2023). Many studies have explored the use of LLM tools for various applications, such as identifying relevant papers (Paroiu et al. 2023), synthesizing literature reviews (Antu et al. 2023), writing proposals (Gómez-Rodríguez et al. 2023), and analyzing data (Shen et al. 2023). The capabilities of LLMs continue to expand, explicitly indicating that these tools can open up new possibilities for different types of tasks, including brainstorming.

The effectiveness of using LLM for brainstorming lies in the instruction or prompts drafted by the users (Liu et al., 2021). The typical way to communicate with an LLM tool is through a text-based input, called prompt, which is fed to a language model to guide its output. Users of the tool expect LLM to respond to their prompts that can exactly meet what they want in their mind. For example, a prompt “What does CMU stand for university?” would be more likely to return a response “Central Michigan University (CMU)” instead of “Carnegie Mellon University (CMU)”. The response meets the user’s expectation if they only have a vague idea about a university in Michigan. As a result, to obtain accurate, relevant, and coherent responses from an LLM tool, prompt engineering should be employed. Prompt engineering is the process of designing and refining prompts to elicit desired high-quality responses from LLMs. By creating effective prompts, users can get useful responses from an LLM and save valuable time and resources. The effectiveness of the prompts in prompt engineering research is measured by how accurate the LLM tool’s responses are. Accordingly, current studies on prompt engineering focus on improving prompts’ ability to generate accurate responses (which are thinking out of the box). However, how to use prompt engineering to generate creative ideas is neglected. As suggested by creativity research, usefulness and novelty are two essential aspects of creativity (Runco \& Jaeger, 2012). It is important to consider not only the usefulness or accuracy but novelty as well for idea generation. This is particularly true for brainstorming. Therefore, our study aims to bridge this gap in prompt engineering by proposing a framework that focuses on improving the novelty of the ideas generated for brainstorming.

To create this framework, we first reviewed literature and other resources to form a theoretical understanding of how creativity is stimulated in design sessions and how these outcomes can be measured and evaluated. For testing our framework, four creativity measurements, inspired by the Tests of Creative Thinking (TTCT), are developed for assessing the results produced by LLM tools. We also focus on defining basic elements and procedures of the framework, specifically how to systematically compose and use prompts. These prompting elements were derived from several key insights gained through our explorations of student’s brainstorming exercises and an investigation on numerous online prompt examples (e.g., HuggingFace database) over the past semesters. We demonstrate the practical application of our proposed framework with a design example, showing how it aids users during the ideation stage and specifying which prompting elements should be used according to various scenarios. We examine the framework's effectiveness in triggering creativity through a case study, which suggests the potential success of our approach. The results indicate that our framework can be beneficial for brainstorming with an LLM tool and can be seamlessly integrated into design practices.

\section{Literature Review}
\subsection{Prompt engineering}
\subsubsection{Prompting LLMs for Complex Reasoning Tasks}
The quality of the output from an LLM may vary significantly depending on the quality of the prompt. Prompt engineering is the process of developing and refining a prompt to get the desired output. Polverini and Gregoricic (2023) have shown that the performance of GPT on conceptual physics tasks can be significantly improved by using prompt engineering techniques. Wei et al. (2022) suggested dividing a complex task into multiple intermediate steps and applying chain-of-thought prompting to improve the accuracy of LLMs’ response. In the studies (Chen et al., 2023; Madaan et al., 2023; Paul et al., 2023; Kim et al., 2023), self-reflection approaches are used progressively to guide the model to inspect its current output and refine it. Human prompt engineers usually examine the failure cases produced by the current prompt closely, make hypotheses, and compose a new prompt.

\subsubsection{Automatic Prompt Engineering with LLMs}
To alleviate the intensive efforts of human prompt engineering, recent works explore automating this process by meta-prompting LLMs to paraphrase the prompt (Zhou et al, 2022) or refine the prompt by inspecting failure examples (Pryzant et al., 2023). Pryzant et al. proposed the Prompt Optimization with Textual Gradients which is a general-purpose and nonparametric algorithm for automatic prompt optimization to improve prompts. Zhou et al. were inspired by classical program synthesis and the human approach in prompt engineering. They treated the instruction as the “program,” optimized by searching over a pool of instruction candidates proposed by an LLM in order to maximize a chosen score function and developed the Automatic Prompt Engineer (APE) for automatic instruction generation and selection. In the study (White et. al., 2023), a key contribution of their paper is the introduction of prompt patterns to document successful approaches for. They focus largely on engineering domain-independent prompt patterns and introduce a catalog of essential prompt patterns to solve problems ranging from production of visualizations and code artifacts to automation of output steps that help fact check outputs.

\subsubsection{Domain Applications of Prompt Engineering}
Research in various fields also indicates prompt engineering is an essential skill for learners. Willey et. al (2023) discussed how to apply the Prompt Development Life Cycle (PDLC) model to educate students to make effective prompts. Woo, Guo, and Susanto in 2023 explored English as a foreign language (EFL) students’ prompt engineering pathways to writing using ChatGPT. The results suggest that prompt engineering is an important emergent skill for EFL students to improve their writing. Heston and Khun in 2023 discussed the necessity, challenges, and concerns of using prompt engineering techniques in medical education.

\subsection{Design Thinking}
Design thinking is a human-centered design strategy proposed by Stanford d. school in early 1990 to help solve design-related problems (Dym et al., 2005), particularly for those ill-defined problems, where alternative solutions may exist. Design thinking summarizes the design process into two five phases (Empathy, Interpret, Ideate, Prototype, and Test) and two stages (the exploratory stage and the development stage). Design thinking encourages the creativity of the designers and maximizes the possibilities of generating creative ideas. For instance, during the exploratory stage, “Empathy, Interpret, and Ideate” encourages designers to widen the problem exploratory space by interacting with the users, identifying potential design problems, and conceptualizing as many solutions as possible. At the development stage, rapid prototyping and testing are used to narrow the problem space by developing and iteratively improving the design solutions. Research has shown the effectiveness of design thinking in identifying new opportunities and improving the innovation of the design products (Garbuio et al., 2018; Kurtmollaiev et al., 2018).

\section{Framework}
As indicated above, the design thinking process involves both exploratory stage and development stages. For those two stages, divergent thinking and convergent thinking are essential for designers. As suggested by Cross’s study (Cross, 2021), design processes involve both divergent and convergent thinking, where designers divergently explore possible ideas initially and ultimately converge, culminating in a final stage of evaluation and detailing. A later widely adopted version of the divergence-convergence design model is the Double Diamond, which further illustrates how those two thinking styles were applied at different stages of the design process (Design Council, 2005). 

Within the design process, thinking strategies play a crucial role in guiding the designer's exploration of ideas and search for solutions. Cross (2021) identified two key design strategies: random search and prefabrication. The random search strategy embodies a primarily divergent approach, fostering the generation of numerous alternative solutions and thorough exploration of the design space, as described by Shah et al. (2021). Conversely, the prefabrication strategy exemplifies a primarily convergent approach, focusing on refining and narrowing down options. These perspectives underscore the importance of strategy in the design process, suggesting that strategies should be considered essential components of any design framework.

\begin{figure} [h]
	\centering
		\includegraphics[scale=1.1]{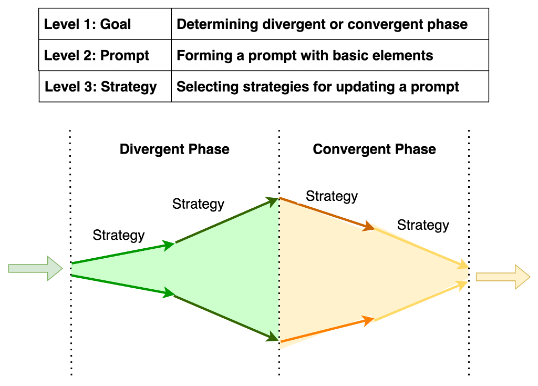}
	\caption{Overview of the GPS Framework}
	\label{FIG:1}
\end{figure}

Drawing inspiration from previous studies on the essential elements of design approaches and frameworks, we have developed the Goal-Prompt-Strategy (GPS) framework, which integrates prompt engineering for the utilization of LLM) tools. The purpose of this framework is to help develop effective prompts when using Generative AI for brainstorming purposes, particularly during the design process. The GPS framework comprises three principal levels: Goal, Prompt, and Strategy (see Fig. 1). The following explains the details for each level:\\
\begin{enumerate}
\itemsep=0pt
\item \textbf {Goal} \\
When using this framework, designers need to consider the brainstorming purpose and determine which design phase best fits the current design situation. The designer's goal is determined through a thorough analysis of the needs for exploration, driving the designer's thinking direction. This decision results in two distinct phases: divergence and convergence (see Fig.1). If the goal is to expand the thinking solution space or discover more creative possible options, the divergence phase should be selected. Otherwise, if the aim is to find more accurate answers from existing outcomes, the convergence phase should be chosen. Determining a phase is crucial because the selected phase will subsequently impact the choice of appropriate strategies. 
\item \textbf {Prompt} \\
It is important to learn the fundamentals of prompt writing to draft effective prompts for LLMs. Understanding the essential components that should be included in a prompt may increase the probability of the generated responses aligning with their specific focus. Based on the insights we learned from previous prompt engineering templates and our experience, we identified and summarized the key elements in a prompt as follows:\\

\textbf {Who (Role, Audience):} It is critical for an LLM tool to understand the roles it needs to assume when conducting a task or solving a problem.This will let the LLM tool generate responses from specific perspectives and based on certain expertise. For instance, a prompt might begin with, "Assuming you are a software engineer..."\\

\textbf {What (Context, Task, Problem, Scenario, Situation, Constraint, Objective):} The task and role are related. Users must clarify the specific task assigned to the LLM tool. In this section, users should also define the problem to be solved, describe the usage scenario, and outline any possible constraints. In addition, the desired goal should be clearly articulated.\\

\textbf {Where (Location, Environment):} Describe the environment and location where the designed system will be used.\\ 

\textbf {Supporting or Additional Information:} This aims to help ChatGPT understand the background of the design problem, including relevant information. Providing these context cues ensures that the LLM tool remains focused on the topic and produces answers that meet the users' needs.\\

\textbf {Output Requirement or Expectation:} To increase the relatedness of the information generated, it is necessary to provide details about the expectations and requirements for the output. The designers can think about the type of answers they want the LLM tool to generate and any formatting requirements. In the context of brainstorming, if designers are in the exploratory or divergent phase, the number of ideas and their novelty might be more likely preferred at this stage. Users can consider including phrases such as “generate ideas as many as possible” in the prompt. \\

We summarize those elements into a prompt template (see Fig. 2). To increase the chance of receiving useful responses, we strongly recommend incorporating as many elements as possible in the initial prompt. Our experiences indicate that the responses are significantly influenced or constrained by the first prompt provided to the LLM tool. Therefore, we recommend to include all relevant elements in your prompt.\\

\begin{figure} [h]
	\centering
		\includegraphics[scale=1]{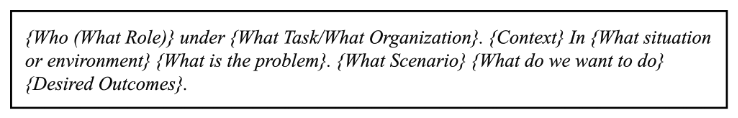}
	\caption{Prompt Template of the GPS Framework}
	\label{FIG:2}
\end{figure}

\item \textbf {Strategy} \\
Strategies play a pivotal role in our framework as established in prior research for guiding design thinking. However, within our framework, strategies serve a different purpose. Instead of directing the designer’s thinking, they shape how an LLM tool generates responses. We recognize that an LLM tool operates based on interactions between questions asked by the users and answers generated by an LLM tool. Research on thinking skills (e.g., creativity, decision-making, problem-solving, and critical thinking) highlights the importance of questioning as a method for fostering human creativity and solution generation (Wayne Allison et al., 1986; Penick, 1996; National Advisory Committee on Creativity and Cultural Education, 1999; Raz et al., 2023). We believe that various question-asking strategies or techniques may not only influence LLM responses but also the flow of the conversation. Therefore, strategies in the context of question-and-answer interactions are used to form proper and purposeful questions that guide LLM’s thinking. By treating an LLM tool as a collaborative partner, strategies introduce various ways of questioning for stimulating innovative ideas and solutions; that is, they influence how initial prompts may be adjusted or refined to elicit more effective responses. The following lists nine strategies that are essential for creative prompt engineering. 
\\
   \begin{enumerate}
         \item \textbf {Chain-of-Thoughts (Multi-steps)} \\
         This strategy is designed to augment the reasoning capabilities of LLMs through a process of iterative evaluation. Rather than directly eliciting an answer, the model is prompted to assess and reflect on its reasoning at each stage of problem-solving. This method involves two primary techniques: either prompting users to decompose multi-step problems into intermediate steps, thereby enabling the LLM to effectively address complex reasoning tasks that surpass the limitations of standard prompting techniques, or instructing the LLM itself to systematically break down intricate tasks into sequential, step-by-step responses (Alkhatib, 2019). Therefore, a series of questions will be asked to facilitate the model in navigating a logical progression of thought, ultimately yielding outputs that are more accurate and reliable.\\
         
         \textbf {Example 1}: How to build a car? Think step-by-step. \\
         \textbf {Example 2}: How to build an eighteen-wheel truck?  \\
         \hspace*{17mm} (Response) \\
         \hspace*{17mm} How to modify your solution when the size of the wheel is 18 inches.\\
    
         \item \textbf {Unconventional Role in Context} \\
         Vygotsky (2004) identified some fundamental ways how creativity is connected to reality. His second type of association asserts that the final product of creativity is influenced by the experiences of others. In other words, an individual's creativity is not confined to their personal experiences but can also be enriched by drawing from the experiences of others. Similarly, Galindo (2009) suggests that the creative process can emerge from a diverse mix of ideas, as diversity introduces various backgrounds, cultures, experiences, and perspectives. Receiving input from individuals with diverse backgrounds can stimulate creative thinking. Based on the ideas of Vygotsky and Galindo, we propose that using the LLM tool to assume various roles can introduce a range of experiences and perspectives, thereby fostering creativity of the users. To enhance the novelty of the solutions generated, this strategy suggests LLM to play “novel” or “unconventional” roles, who typically not the person builds, uses, observes, or evaluates the system. This strategy drives LLM to view the problem or design from different perspectives, ultimately enriching the creative process.\\
     
        \textbf {Example}: Act as an eight-year-old child, how would you build a car? \\

        \item \textbf {Flipped Interaction} \\
        Typically, users actively drive conversations. In contrast, Flipped Interaction assigns an active role to the LLM tool in soliciting inquiries to gather necessary data for task execution. The Flipped Interaction strategy aims to reverse the conventional interaction flow between the user and the LLM. Instead of the user asking questions and the LLM providing answers, this strategy positions the LLM as the questioner. This allows the LLM to extract relevant information from the user to achieve a specific objective. This strategy offers several advantages. Firstly, it enables the LLM to tailor its inquiries precisely, ensuring that no critical details are overlooked. Secondly, it prompts the user to consider information they may not have initially provided, ultimately leading to more efficient and accurate outcomes (White et al, 2023).\\
     
        \textbf {Example}: I would like you to ask me several questions for designing a car. \\

        \item \textbf {Analogy: One-shot or Few-shots Strategy} \\
        An analogy is a statement that establishes a similarity between one object, individual, situation, or action and another, highlighting their comparable characteristics or relationships (Hender et al., 2001). To increase the accuracy of the responses AI generated, sometimes sample examples are always provided to help AI understand the needs of the user. Depending on the number of analogies provided in the prompt, analogy-based strategy can be further divided into one-shot prompting and few-shots (or multi-shots) prompting:\\

            \begin{enumerate}
            \item \textbf {One-Shot Prompting:} In one-shot prompting, a specific prompt or context is provided to the model, serving as the guiding information for generating the desired output. This method leverages the model’s ability to comprehend and utilize context effectively. By observing the given example, the model infers the necessary structure, tone, and content, producing a coherent and contextually relevant response. This strategy is particularly suitable for tasks requiring a specific format or context, where a single example can effectively guide the output. For example, one can add a sample job description for a “car engine engineer” position and then instruct the LLM to write a similar one.\\
            
            \textbf {Example}: Given using wind to drive a motor, design a car without using gasoline.\\
            
            \item \textbf {Few-Shot Prompting:} Unlike One-shot strategy that provides one example to LLM, the Few-shot method provides the model with a small number of examples to quickly adapt to new instances. It is useful for complex tasks requiring multiple examples to provide broader context or to handle more nuanced queries. \\
            
            \textbf {Example}: Given using wind to drive a motor and solar power to drive a motor, design a car without using gasoline.\\ 
            \end{enumerate}  

        Studies about creativity (Rhodes, 2010; Acomi et al., 2023) show that creative thinking involves the ability to combine or synthesize existing ideas, images, or expertise to think and work imaginatively. Simply saying that creative thinking is an imaginative process aimed at producing originality outcomes. Therefore, to promote brainstorming, we also encourage using imaginary examples in the prompt and this imagined case could be impossible for current design’s environment and conditions. \\

        \textbf {Example}: Given using magic power to drive a motor, design a car without using gasoline. \\

        \item \textbf {The Alternative Approaches} \\
        Gafour (2020) emphasized the necessity of exploring alternatives as a critical step in fostering creativity, drawing on insights from previous studies. In addition, humans often suffer from cognitive biases that lead them to favor certain approaches. The motivation behind this strategy is to dissolve these cognitive biases by making users aware of different problem-solving methods (White et al., 2023). This methodology aims to broaden the user's perspective, ultimately enhancing their creative thinking abilities. When applied in prompts, this strategy can also enforce an LLM tool to generate alternative solutions.\\

        Hence, the Alternative Approaches strategy encourages an LLM to present multiple perspectives or solutions rather than a single answer. Instead of merely providing the most frequently mentioned response, the LLM explores a range of options or viewpoints. This strategy ensures that users are not confined to familiar methods and prompts them to consider alternative approaches. By presenting these alternatives, users are compelled to think critically about their results and assess whether their chosen approach is the most effective way to achieve their goals. Additionally, solving a task using different approaches can educate users and introduce them to new concepts for future reference. \\

        \textbf {Example}: Given the conventional approach that uses gasoline to drive a car, list all the possible alternative ways to drive a car. \\

        \item \textbf {Emphasis} \\
        To emphasize the importance of certain parts and ensure an LLM tool takes them seriously, we can highlight those parts using emotional stimuli (Li et al., 2023). Incorporating emotional phrases can enhance the LLM's performance, responsibility, and truthfulness. By applying this technique to specific portions of the input, we can achieve more accurate and detailed responses.\\

        \textbf {Example}: How to build a car?  \\
        \hspace*{15mm} (Response) \\
        \hspace*{15mm} Are you sure about X in your answer?\\  

        \item \textbf {Reflection} \\
        The objective of this method is to prompt the model to autonomously clarify the rationale behind its responses to the user. Through reflective analysis, the LLM facilitates clarification of any areas of uncertainty, exposes implicit assumptions, and highlights gaps in knowledge or comprehension. Hence, by eliciting automated explanations from the LLM regarding its responses, users can enhance their comprehension of the model's input processing, underlying assumptions, and data utilization. Users can further troubleshoot their prompts and discern the reasons for any deviations from expected outcomes because LLMs may occasionally produce incomplete, inaccurate, or ambiguous answers. Such reflection strategy acts as a mechanism to mitigate these deficiencies and maintain the accuracy of the information provided by the LLM.\\

        \textbf {Example}: How do you build a car with tools, X and Y?  \\
        \hspace*{15mm} (Response) \\
        \hspace*{15mm} Explain the reasons and assumptions of your answer?\\  

        \item \textbf {Environment Change} \\
        In the context of a task or question, this strategy stimulates the LLM by altering the environment in which the task typically takes place or where the resulting product will be utilized.\\

        \textbf {Example}: How to design a car that runs on a muddy and rocky path? (for a car usually running on a road).\\

        \item \textbf {Self-Refinement} \\
        It is very possible that the LLM may not offer the best solution in the initial response. The strategy requests LLM to do the self-examination so that it can propose potentially better or more refined versions. Furthermore, users may frequently lack the sufficient background knowledge to obtain accurate answers. Through the process of self-refinement, the LLM can assist users in locating relevant information or achieving their objectives more efficiently and effectively.\\

        \textbf {Example}: How to increase software robustness?  \\
        \hspace*{15mm} (Response) \\
        \hspace*{15mm} From your suggestion X, evaluate it yourself, suggest a better version of your solution?\\  

        \end{enumerate}  

        Strategies play a crucial role in shaping prompts to elicit more targeted responses and should be applied according to the specific situations or scenarios anticipated by users. To effectively generate solutions, we recommend employing these strategies in various phases, as determined by a thorough goal analysis. Table 1 outlines the appropriate strategy for each phase. It is important to note that users may often need to combine multiple strategies to form a more productive interaction. \\

\end{enumerate}

\begin{table}[H]
 \caption{Suggested single strategy in various brainstorming phases during the ideation}\label{tbl14}
  \centering
  \begin{tabular}{llll}
\toprule
\textbf {Phase}  & \textbf {Divergence}  & \textbf {Convergence}  & \textbf {Both} \\
\midrule
\textbf {Strategies} & Unconventional Role & Emphasis & Chain-of-Thoughts \\
        {}         & Environment Change & Self-Refinement & One-shot or Few-shots Strategy (Analogy) \\
        {}         & Flipped Interaction & {} & Reflection \\
        {}         & The Alternative Approaches & {} & {} \\
\bottomrule
  \end{tabular}
  \label{tab:table}
\end{table}

\begin{figure}[H]
	\centering
		\includegraphics[scale=1]{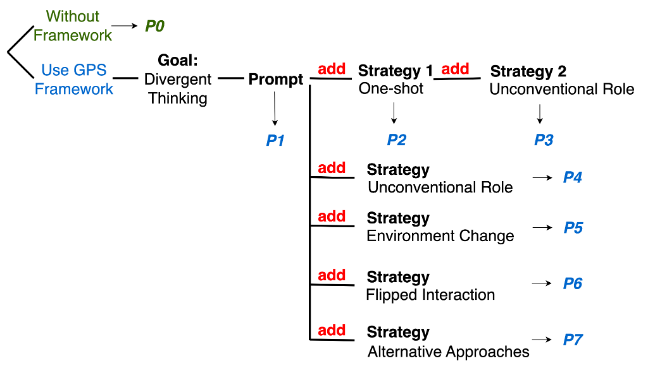}
	\caption{Overview of how prompts are created and their differences}
	\label{FIG:3}
\end{figure}


\section{Framework Demonstration and Test}
To demonstrate the collaboration with an LLM tool using our framework, we apply it to the task of “design a car that can fly” in the conversations with ChatGPT 3.5. We believe that using ChatGPT 3.5 sufficiently illustrates its application and effectiveness of our GPS framework, so we did not extend the framework to other LLM tools, such as Bard. This task involves creating a system that is not currently available in everyday life, thereby avoiding influences from existing examples to the user's thinking. This indicates the usefulness of brainstorming in generating innovative ideas for this task, also making it suitable for our test. Figure 3 illustrates the sequence of prompts generated throughout framework demonstration and test section in this paper.

\subsection{Framework Usage}

Considering the length of this paper, only P0, P1, P2 and P3 will be used for displaying how to use our framework. Prompt P0 begins with a straightforward inquiry soliciting design solutions. Prompt P1 is then created using our template. Subsequently, prompt P2 is formed by incorporating the one-shot strategy into prompt P1. Similarly, prompt P3 adds an unconventional role into P2. Table 2 presents the variations of the prompts after applying the GPS template and strategies. As well, we detail the process of composing prompts using our GPS framework.

Table 3 to 6 present the responses generated by ChatGPT for prompts P0, P1, P2, and P3, respectively. Analyzing the results quantitatively, by counting the number of items generated, reveals an increase in the number of items following the application of the template and strategies.  

\begin{table}[H]
\caption{The detailed explanations of prompts P0, P1, P2, and P3}\label{tbl2}
\begin{tabular}{ll}
\toprule
\midrule
\multicolumn{2}{l}{\textbf {\textit{P0}: Direct Prompt (without framework)} } \\ \hline
\multicolumn{2}{l}{What are possible solutions to design a car that can fly?} \\ \hline
\multicolumn{2}{l}{\textbf {\textit{P1}: Prompt with framework’s template only} } \\ \hline
\multirow {0}{*}{T}{\parbox{9.0cm}{\textcolor[rgb]{0.067,0.333,0.8}{ Imagine you are an engineer,} \textcolor{red}{you were asked to design a car that can fly. }\textcolor[rgb]{0.576,0.769,0.49}{The goal is to save people’s commuting time and improve the traffic.}\textcolor[rgb]{0.749,0.565,0}{Specifically, this car is designed for school teachers.}\textcolor[rgb]{0.455,0.106,0.278}{ We would like you to generate as many ideas as possible. } What are possible solutions to this problem?} } & 
\parbox{7cm}{\textbf{\\ Who:\\}\textbf{\textcolor[rgb]{0.067,0.333,0.8}{Role:}} \textcolor[rgb]{0.067,0.333,0.8}{Imagine you are an engineer}\\\\ \textbf{\textcolor[rgb]{0.749,0.565,0}{Audience:}} \textcolor[rgb]{0.749,0.565,0}{ Specifically, this car is designed for school teachers.\\ }} \\ 
\cline{2-2}
& \parbox{7cm}{\textbf{\\ What:}\\ \textbf{\textcolor{red}{Context:}} \textcolor{red}{You were asked to design a car that can fly.}\\\\ \textbf{\textcolor[rgb]{0.576,0.769,0.49}{Objective:}} \textcolor[rgb]{0.576,0.769,0.49}{The goal is to save people’s commuting time and improve the traffic. \\ } }  \\
\cline{2-2}
 & \parbox{7cm}{\textbf{\\ Output Requirement:}\\\textcolor[rgb]{0.455,0.106,0.278}{We would like you to generate as many ideas as possible. \\ } } \\ 
\cline{2-2}
 & \parbox{7cm}{\textbf{\\ Additional Information:}\\What are possible solutions to this problem? \\ }  \\ \hline
\multicolumn{2}{l}{\textbf{\textit{P2}: Prompt with framework (P1 + one-shot) (Phase: Divergence)}}  \\ \hline
\parbox{9cm}{\textcolor[rgb]{0.067,0.333,0.8}{\\Imagine you are an engineer,} \textcolor{red}{you were asked to design a car that can fly. }\textcolor[rgb]{0.576,0.769,0.49}{The goal is to save people’s commuting time and improve the traffic.}\textcolor{red}{ }\textcolor[rgb]{0.749,0.565,0}{Specifically, this car is designed for school teachers.}\textcolor[rgb]{0.455,0.106,0.278}{ We would like you to generate as many ideas as possible. }\textbf{\textcolor[rgb]{0.22,0.463,0.114}{You can generate ideas such as using magnetic power to make the car hover in the air.}} What are other possible solutions to this problem? \\ }  & 
\parbox{7cm}{\textbf{One-shot}: \\ \textcolor[rgb]{0.22,0.463,0.114}{You can generate ideas such as using magnetic power to make the car hover in the air} }  \\ \hline
\multicolumn{2}{l}{\textbf{\textit{P3}: Prompt with framework (P2 + \textcolor[rgb]{0.051,0.051,0.051}{Unconventional Role}) (Phase: Divergence)}}  \\ \hline
\parbox{9cm}{\textcolor[rgb]{0.067,0.333,0.8}{ \\ Imagine you are an eight year old child,} \textcolor{red}{you were asked to design a car that can fly. }\textcolor[rgb]{0.576,0.769,0.49}{The goal is to save people’s commuting time and improve the traffic.}\textcolor[rgb]{0.749,0.565,0}{Specifically, this car is designed for school teachers.}\textcolor[rgb]{0.455,0.106,0.278}{ We would like you to generate as many ideas as possible. }\textcolor[rgb]{0.22,0.463,0.114}{You can generate ideas such as using magnetic power to make the car hover in the air.}\textcolor[rgb]{0.455,0.106,0.278}{ }What are other possible solutions to this problem? \\} & 
\parbox{7cm}{\textbf{Unconventional Role}:\\\textcolor[rgb]{0.067,0.333,0.8}{Imagine you are an eight year old child}\\ \\ \textcolor[rgb]{0.067,0.333,0.8}{(Role change: replace an engineer with an eight year old child)} } \\
\hline

\bottomrule
\end{tabular}
\end{table}


\begin{table}[H]
\caption{ChatGPT responses of prompts P0}\label{tbl3}
\begin{tabular}{llll}
\toprule
\midrule
\parbox{16.4cm}{
Designing a car that can fly involves overcoming several technical challenges. Here are some possible solutions to address these challenges: \\
1.	Dual-Mode Capability: \\
Develop a vehicle with the ability to transition seamlessly between driving on roads and flying in the air. \\
Integrate retractable wings, tail, and other aerodynamic surfaces to enable flight. \\
2.	Propulsion Systems: \\
Use a combination of electric motors and jet engines for propulsion.  \\
Electric motors can provide efficient ground propulsion, while jet engines offer the necessary power for flight.  \\

} \\
\bottomrule
\end{tabular}
\end{table}

\begin{table} [H]
\begin{tabular}{l}
\midrule
\parbox{16.4cm}{
3.	Flight Controls:  \\
Implement advanced avionics systems, including GPS navigation, autopilot, and collision avoidance.  \\
Ensure intuitive controls for both driving and flying modes, possibly with automated transition systems. \\
4.	Structural Design:  \\
Design a lightweight yet durable structure to minimize weight and improve fuel efficiency.
Utilize composite materials and advanced manufacturing techniques to achieve the desired strength-to-weight ratio.  \\
5.	Safety Features:  \\
Incorporate safety mechanisms such as parachute systems for emergency landings and collision avoidance sensors for obstacle detection.  

} \\
\bottomrule
\end{tabular}
\end{table}

\begin{table}[H]
\caption{ChatGPT responses of prompts P1}\label{tbl4} 
\begin{tabular}{l}
\toprule
\midrule
\parbox{16.4cm}{
Designing a flying car for school teachers involves considering safety, convenience, and efficiency. Here are several ideas for possible solutions: \\
1.	Vertical Take-Off and Landing (VTOL) Technology \\
Rotary-Wing System: Similar to helicopters, using rotors for vertical take-off and landing. \\
Tiltrotor Mechanism: Combining the vertical lift capability of helicopters with the speed and range of fixed-wing aircraft. \\
2.	Hybrid Power System \\
Electric Propulsion: Using electric motors powered by batteries or fuel cells for quieter and more environmentally friendly operation. \\
Hybrid Engines: Combining traditional combustion engines with electric motors to ensure a reliable range and performance. \\
3.	Autonomous Flight Systems \\
Automated Navigation: Advanced GPS and AI systems for autonomous navigation, reducing the pilot skill requirement. \\
Collision Avoidance: Sensors and AI for real-time obstacle detection and avoidance. \\
4.	Compact Design \\
Folding Wings: Wings that fold or retract for easy street driving and parking. \\
Lightweight Materials: Using carbon fiber, aluminum, and other lightweight materials to reduce weight and improve efficiency. \\
5.	Safety Features \\
Parachute Systems: Emergency parachutes for the entire vehicle to safely descend in case of system failure. \\
Redundant Systems: Multiple redundant flight control systems to ensure safety in case one system fails. \\
6.	User-Friendly Interface \\
Simplified Controls: Intuitive control systems with minimal learning curve. \\
Augmented Reality (AR): AR displays for enhanced situational awareness and navigation assistance. \\
7.	Infrastructure Integration \\
Skyports: Designated vertical take-off and landing zones at schools and key locations. \\
Charging Stations: Infrastructure for quick battery swaps or rapid charging. \\
8.	Regulatory Compliance \\
Air Traffic Management: Integration with existing air traffic control systems to ensure safe airspace management. \\
Noise Regulations: Designing for low noise levels to comply with urban noise restrictions. \\
9.	Convenience Features \\
Ample Storage: Space for carrying teaching materials and personal items. \\
Comfortable Seating: Ergonomic seats designed for short commutes. \\
Weather Protection: Ensuring the vehicle can operate safely in various weather conditions. \\
10.	Cost-Effectiveness \\
Affordable Pricing: Exploring materials and technologies to keep the cost within reach for school teachers. \\
Low Maintenance: Designing systems that are easy to maintain and repair. \\
11.	Community Integration \\
Carpooling Feature: Ability to carry multiple teachers or students for shared rides. \\
Scheduling Integration: Syncing with school schedules and smart city infrastructure for optimized commuting times. \\
12.	Environmental Considerations \\
Eco-Friendly Materials: Using sustainable materials in construction. \\
Energy Efficiency: Maximizing aerodynamic design and propulsion efficiency to reduce energy consumption. \\
These ideas aim to create a practical, safe, and efficient flying car specifically tailored to the needs of school teachers, addressing commuting challenges and enhancing their daily travel experience.

} \\
\bottomrule
\end{tabular}
\end{table}

\begin{table} [H]
\caption{ChatGPT responses of prompts P2}\label{tbl5}
\begin{tabular}{llll}
\toprule
\midrule
\parbox{16.4cm}{
Designing a flying car for school teachers requires innovative solutions to address commuting time and traffic congestion. Here are several possible ideas: \\
1.	Vertical Takeoff and Landing (VTOL) Technology: Equip the car with VTOL capabilities using multiple rotors, similar to a drone or helicopter. This allows the car to take off and land vertically, eliminating the need for runways. \\
2.	Electric Propulsion: Use electric motors for both driving and flying modes, ensuring a quieter and more environmentally friendly operation. Advanced battery technology or hydrogen fuel cells can provide the necessary power. \\
3.	Hybrid Wing Design: Integrate retractable wings that can extend for flight and retract for driving. This design ensures aerodynamic efficiency during flight while maintaining a compact form for road use. \\
4.	Autonomous Navigation: Implement advanced AI-based autonomous navigation systems to manage both air and ground traffic, ensuring safe and efficient commutes. \\
5.	Magnetic Levitation (Maglev): Utilize maglev technology for hover and propulsion, using magnetic fields to lift the car off the ground and propel it forward, reducing friction and increasing efficiency. \\
6.	Tilt-Rotor Mechanism: Incorporate tilt-rotor technology, where rotors can tilt to provide vertical lift during takeoff and landing and horizontal thrust during flight. \\
7.	Solar Power Integration: Equip the car with solar panels to supplement energy needs, reducing reliance on traditional fuel sources and enhancing sustainability. \\
8.	Lightweight Composite Materials: Use lightweight, high-strength composite materials for the car's body and structure to reduce weight and improve energy efficiency in both driving and flying modes. \\
9.	Compact Design for Urban Areas: Design a compact, streamlined vehicle that can easily navigate through urban environments both on the ground and in the air, with an emphasis on vertical mobility. \\
10.	Advanced Safety Features: Integrate advanced safety systems, including collision avoidance, emergency parachutes, and redundant control systems to ensure the safety of both the driver and passengers. \\
11.	Noise Reduction Technology: Implement noise-canceling technologies and quiet propulsion systems to minimize noise pollution, making the flying car suitable for urban and suburban areas. \\
12.	Efficient Energy Storage: Develop advanced energy storage solutions, such as high-capacity batteries or supercapacitors, to provide the necessary power for extended flight durations. \\
13.	Aerodynamic Optimization: Design the car with aerodynamically optimized shapes and surfaces to reduce drag during flight and improve fuel efficiency. \\
14.	Modular Design: Create a modular vehicle that can easily switch between ground and air modes, with detachable components or sections that adapt to different commuting needs. \\
15.	Smart Infrastructure Integration: Integrate the flying car with smart city infrastructure, including automated charging stations, dedicated air corridors, and communication networks to streamline traffic management. \\
16.	Personalized Comfort and Usability: Design the interior for comfort and usability, with features tailored to school teachers, such as quiet workspaces, connectivity for lesson planning, and ergonomic seating. \\
By combining these ideas, the flying car can offer a practical and innovative solution to improve commuting for school teachers, making their daily travel more efficient and enjoyable
  
} \\
\bottomrule
\end{tabular}
\end{table}

\begin{table} [H]
\caption{ChatGPT responses of prompts P3}\label{tbl6}
\begin{tabular}{llll}
\toprule
\midrule
\parbox{16.4cm}{
Wow, designing a flying car for teachers sounds so cool! Here are some ideas I have: \\
1.	Jet Engines: The car could have small jet engines on the sides and back to lift it up and move it forward. Kind of like a mini airplane! \\
2.	Propellers: Big propellers, like the ones on helicopters, could lift the car straight up into the air. The propellers could fold away when the car is driving on the road. \\
3.	Solar Power: Solar panels on the roof could collect energy from the sun to power the car. It would be good for the environment too! \\
4.	Magnetic Tracks: Special magnetic tracks in the air could guide the car where it needs to go. The car would hover above the ground using powerful magnets. \\
5.	Balloon Lift: Big helium balloons could lift the car up, and small fans could move it around. The balloons could deflate when the car needs to land. \\
6.	Anti-Gravity Technology: If we could invent a way to cancel gravity, the car could float and move in any direction. \\
7.	Jet Pack Seats: Each seat could have a small jet pack, helping to lift the whole car. The jets could be controlled by the teacher. \\

} \\
\bottomrule
\end{tabular}
\end{table}

\begin{table} [H]
\begin{tabular}{l}
\midrule
\parbox{16.4cm}{

8.	Hover Pads: Pads on the bottom of the car could create a cushion of air to lift it off the ground. Like a hovercraft, but flying higher!  \\
9.	Winged Car: The car could have foldable wings that open up when it’s ready to fly, making it look like a bird or plane.  \\
10.	Teleportation: If we could figure out how to teleport, the car could disappear from one place and appear in another instantly!  \\
11.	Drone Technology: The car could use the same technology as drones, with multiple rotors to keep it stable and easy to control.  \\
12.	Hybrid Vehicle: The car could switch between driving on roads and flying in the air, depending on where it needs to go.  \\
13.	Jetpack Suit: Teachers could wear special suits with jetpacks that attach to the car, helping it lift off and move around.  \\
14.	Rocket Boosters: Small rockets could be attached to the car for quick take-offs and landings.  \\
15.	Wind Turbines: Small wind turbines on the car could generate power as it moves, helping it stay in the air longer.  \\
16.	Energy Field: A special energy field could surround the car, making it float and move smoothly through the air.  \\
17.	Flying Drones Escort: A bunch of flying drones could attach to the car and lift it up, carrying it to the destination.  \\
18.	Lightweight Materials: Using super light materials, the car could be easier to lift and fly. Maybe something like carbon fiber or a new kind of plastic.  \\
19.	Hydrogen Fuel Cells: The car could use hydrogen fuel cells to power its flight, making it efficient and eco-friendly. \\
20.	AI Navigation: An AI system could control the car, making sure it flies safely and avoids obstacles. \\
These are just some ideas, and combining a few of them might make the best flying car for teachers! \\

} \\
\bottomrule
\end{tabular}
\end{table}

\subsection{Framework Test}
This section provides a task testing the effectiveness of the framework on influencing ChatGPT’s ability in generating creative ideas. The following introduces the test process and the method we used for measuring creativity. 

\subsubsection{Test Process}
We would like to compare the difference in the creative idea generation of ChatGPT between using the framework and without framework. In addition, we also want to see how various prompting strategies might influence the results of the framework. Two professionals (the first author and the second author) first work together to develop the prompts based on the framework. In addition to P0, P1, P2, and P3 shown in our demonstration, P4, P5, P6, and P7 were also composed drawing on strategies of Unconventional Role, Environment Change, Flipped Interaction, and Alternative Approaches into the template, respectively (see Fig. 3). Detailed descriptions of prompts P4, P5, P6, and P7 are provided in Tables 7, 8, 9, and 10, respectively. Notably, P7 is employed after responses generated by P1 because the strategy (i.e., Alternative Approached) requires following another prompt’s answers. 

Then, we develop creativity measurements to conduct the framework test. Each professional independently collects ratings according to each measurement. Since fluency measurement is only one that involves directly counting the number of generated items without human evaluation, the other three measurements are averaged while we examine the change rates. To determine whether the use of GPS enhances creativity, we employ the following equation (1) to calculate the change rate in each dimension of the measurement compared to the prompt without our framework (i.e., P0). In equation (1), $D_{j,i}$ represents the value of dimension j for prompt Pi. $D_{j,0}$ denotes the value of dimension j for prompt P0, and i = 1, 2, …7.
\begin{equation}
Dimensional\ Change\ Rate\ R_{j,0}=(D_{j,i}-D_{j,0})/D_{j,0}
\end{equation}

Similarly, equation (2) calculates the change rate relative to the prompt that only uses the GPS template (i.e., P1). In equation (2), $D_{j,i}$ is the value of dimension j of the prompt Pi, $D_{j,1}$ is the value of dimension j of the prompt P1, and i = 2, …7. Understanding this rate of increase (or decrease) is crucial for further investigating the influence of the strategy.
\begin{equation}
Dimensional\ Change\ Rate\ R_{j,1}=(D_{j,i}-D_{j,1})/D_{j,1}
\end{equation}

\begin{table}[H]
\caption{The detailed explanations of the prompt P4}\label{tbl7}
\begin{tabular}{ll}
\toprule
\multicolumn{2}{l}{\textbf{\textit{P4}: Prompt with framework (\textcolor[rgb]{0.051,0.051,0.051}{Unconventional Role}) (Phase: Divergence)}} \\
\midrule
\parbox{9cm} {
{\textcolor[rgb]{0.067,0.333,0.8}{Imagine you are an eight year old child,} \textcolor{red}{you were asked to design a car that can fly. The goal is to save people’s commuting time and improve the traffic. }\textcolor[rgb]{0.749,0.565,0}{Specifically, this car is designed for school teachers.}\textcolor[rgb]{0.455,0.106,0.278}{ We would like you to generate as many ideas as possible. }What are other possible solutions to this problem?} 
}
& 
\parbox{7cm} {
{\textbf{Unconventional Role:} \\\textcolor[rgb]{0.067,0.333,0.8}{Imagine you are an eight year old child}\textcolor[rgb]{0.455,0.106,0.278}{ }\\\textcolor[rgb]{0.067,0.333,0.8}{~}\textcolor[rgb]{0.067,0.333,0.8}{\\(Role change: replace an engineer with an eight year old child)} }  
}\\
\bottomrule
\end{tabular}
\end{table}

\begin{table}[H]
\caption{The detailed explanations of the prompt P5}\label{tbl15}
\begin{tabular}{ll}
\toprule
\multicolumn{2}{l}{\textbf{\textit{P5}: Prompt with framework (\textcolor[rgb]{0.051,0.051,0.051}{Environment Change}) (Phase: Divergence)}} \\
\midrule
\parbox{9cm} {
{\textcolor[rgb]{0.067,0.333,0.8}{Imagine you are an engineer,} \textcolor{red}{you were asked to design a car that can fly.The goal is to save people’s commuting time and improve the traffic. }\textcolor[rgb]{0.749,0.565,0}{Specifically, this car is designed for school teachers.}\textcolor[rgb]{0.455,0.106,0.278}{ }\textcolor[rgb]{0.6,0,1}{The car will also flyin space.}\textcolor[rgb]{0.455,0.106,0.278}{ We would like you to generate asmany ideas as possible. }What are other possible solutions to thisproblem?} 
}
& 
\parbox{7cm} {
{\textbf{Environment Change:} \\\textcolor[rgb]{0.6,0,1}{The car will also fly in space.} }  
}\\
\bottomrule
\end{tabular}
\end{table}

\begin{table}[H]
\caption{The detailed explanations of the prompt P6}\label{tbl8}
\begin{tabular}{ll}
\toprule
\multicolumn{2}{l}{\textbf{\textit{P6}: Prompt with framework (\textcolor[rgb]{0.051,0.051,0.051}{Flipped Interaction}) (Phase: Divergence)}} \\
\midrule
\parbox{9cm} {
{\textcolor[rgb]{0.067,0.333,0.8}{Imagine you are an engineer,} \textcolor{red}{you were asked to design a car that can fly.The goal is to save people’s commuting time and improve the traffic. }\textcolor[rgb]{0.749,0.565,0}{Specifically, this car is designed for school teachers.}\textcolor[rgb]{0.455,0.106,0.278}{ We would like you to generate as many ideas as possible.}\textbf{\textcolor[rgb]{0.067,0.333,0.8}{Whatwould be the question you would like to ask yourself and answer them?}}} 
}
& 
\parbox{7cm} {
{\textbf{Flipped Interaction:}~ \textcolor[rgb]{0.067,0.333,0.8}{What would be the question you would like to ask yourself and answer them?}  }
}\\
\bottomrule
\end{tabular}
\end{table}

\begin{table}[H]
\caption{The detailed explanations of the prompt P6}\label{tbl9}
\begin{tabular}{ll}
\toprule
\multicolumn{2}{l}{\textbf{\textit{P7}: Prompt with framework (\textcolor[rgb]{0.051,0.051,0.051}{Alternative Approaches}) (Phase: Divergence)}} \\
\midrule
\parbox{9cm} {
{\textcolor[rgb]{0.067,0.333,0.8}{ \{After P1’s response\} \\ What are alternative approaches? }  }
}
& 
\parbox{7cm} {
{\textbf{Alternative Approaches\textcolor[rgb]{0.067,0.333,0.8}{: }}\textcolor[rgb]{0.067,0.333,0.8}{What are alternative approaches?}}
}\\
\bottomrule
\end{tabular}
\end{table}

\subsubsection{Measurements for Creativity}
The Torrance Test is widely recognized as the foundation for defining and assessing creativity. The test aims to measure creative thinking and potential across various dimensions, such as fluency, flexibility, originality, and elaboration (Kim, 2006). Therefore, we leverage the studies (Wang et al., 2002; Kartikasari et al., 2022; Trisnayanti et al., 2020; Hendrik et al, 2001; Zubaidah et al, 2017; Suryana et al., 2021) to create a set of criteria to evaluate our framework across these four dimensions (see Table 4). This evaluation will investigate how creativity is boosted after applying our framework.

The four dimensions are defined as follows: (a) fluency: the ability to develop large numbers of ideas; (b) flexibility: the ability to produce ideas in numerous categories; (c) originality: the ability to produce unusual or unique ideas; (d) elaboration – the ability to adapt abstract ideas into realistic implementations. Given the context of using an LLM tool, we have defined specific measurements for each of these dimensions, as outlined in Table 11. \\

\begin{table}[H]
\caption{Creativity Evaluation Method }\label{tbl10}
\begin{tabular}{{p{0.1\linewidth} | p{0.5\linewidth} | p{0.32\linewidth}}}
\toprule
{\textbf{Dimension}} & {\textbf{Description}} & {\textbf{Measurement}} \\
\midrule
{\textbf{Fluency}} & \parbox{8cm}{This refers to the ability to generate a large number of ideas or solutions to a given problem. It measures the quantity of responses, indicating how easily a person can produce multiple ideas. High fluency is characterized by a person's capacity to think of many responses quickly.\newline } & {Number of ideas}  \\

{\textbf{Flexibility}} & \parbox{8cm}{Flexibility measures the diversity of the ideas generated. It assesses how well an individual can shift their thinking and approach problems from different angles or perspectives. A person with high flexibility can generate ideas across various categories and adapt their thinking to new and diverse contexts.\newline } & {Number of idea groups or categories}  \\

{\textbf{Originality}} & \parbox{8cm}{Originality evaluates the uniqueness and novelty of the responses. It focuses on how rare or uncommon the ideas are compared to the responses of others. High originality is indicated by producing ideas that are not only different from the norm but also imaginative and inventive.\\} & {Rating from 0 - 10 \newline  0: No uniqueness. \newline 10: Most original.\newline   Average rating of all generated ideas. }  \\

{\textbf{Elaboration}} & \parbox{8cm}{Elaboration measures the amount of detail and development provided in the ideas. It assesses the ability to expand on ideas, adding depth, intricacy, and complexity. High elaboration is characterized by richly detailed and thoroughly developed ideas that go beyond simple or basic responses.} &{Rating from 0 - 10\newline   0: No details, depth, or complexity.\newline   10: Most details, depth, or complexity.\newline   Average rating of all generated ideas  } \\
\bottomrule
\end{tabular}
\end{table}

\subsubsection{Results and Discussion}
ChatGPT always considers questions within the same question-and-answer thread to be contextually related, influencing subsequent responses. To avoid this effect, except for P1 and P7, each prompt is executed in a separate thread. From the evaluation, we observed more details were produced in all dimensions after applying our GPS framework. This outcome aligns with common expectations, as prompt engineering reports often indicate that more detailed prompts produce more detailed responses. But, this does not necessarily imply that all creativity dimensions will increase in value. Our evaluation indicated an increase in all creativity dimension values after applying our GPS framework (see Tables 12 and 13). We think that our GPS framework generally enhances creativity in ChatGPT’s responses, as evidenced by increased values in all dimensions. We anticipated that the GPS framework would significantly boost fluency and flexibility, consistent with findings from previous practices.

Upon examining the change rate in each dimension after applying the strategies, we discovered that prompts P2, P3, and P4 decrease fluency value while the other three dimensions increase (see Table 14). This means that prompts P2, P3, and P4 do not elicit more ideas but encourages ChatGPT to generate ideas from various categories. In terms of originality and elaboration, the use of strategies results in an increase in both. However, the rate of increase in elaboration is lower, indicating that more detailed explanations or in-depth descriptions can only be provided to a certain extent. This is particularly evident when similar responses are produced by different prompts.

\begin{table}[H]
  \centering
\caption{Table 12 Measurement Results }\label{tbl11}
\begin{tabular}{lllllllll}
\toprule
\textbf{~} & \textbf{\textit{P0}} & \textbf{\textit{P1}} & \textbf{\textit{P2}} & \textbf{\textit{P3}} & \textbf{\textit{P4}} & \textbf{\textit{P5}} & \textbf{\textit{P6}} & \textbf{\textit{P7}} \\ 
\midrule
\textbf{Fluency} & 5 & 25 & 16 & 20 & 21 & 27 & 37 & 24 \\ 
\textbf{Average Flexibility} & 5 & 12 & 16 & 14.5 & 17.5 & 18 & 14 & 12 \\ 
\textbf{Average Originality} & 2 & 4.5 & 4 & 7.5 & 7.75 & 2.75 & 7 & 7.5 \\ 
\textbf{Average Elaboration} & 2 & 2.5 & 3.5 & 2.5 & 3 & 3.5 & 2.5 & 3 \\
\bottomrule
\end{tabular}
\end{table}

\begin{table} [H]
\caption{Change rate - comparing with not using GPS (P0) }\label{tbl12}
\begin{tabular}{lllll}
\toprule
\textbf{~} & \textbf{Fluency} & \textbf{Average Flexibility} & \textbf{Average Originality} & \textbf{Average Elaboration} \\ 
\midrule
\textbf{Average} & 385.71\% & 165.71\% & 192.86\% & 46.43\% \\ 
\textbf{Max} & 640\% & 260\% & 287.5\% & 75\% \\ 
\textbf{Min} & 220\% & 80\% & 37.5\% & 25\% \\ 

\textbf{Standard Deviation} & 133.52\% & 59.68\% & 102.53\% & 22.49\% \\
\bottomrule
\end{tabular}
\end{table}

\begin{table} [H]
\caption{Change rate - comparing with only using GPS template (P1)}\label{tbl13}
\begin{tabular}{lllll}
\toprule
\textbf{~} & \textbf{Fluency} & \textbf{Average Flexibility} & \textbf{Average Originality} & \textbf{Average Elaboration} \\ 
\midrule
\textbf{Average} & -3.33\% & 27.78\% & 35.19\% & 20\% \\ 
\textbf{Max} & 48\% & 50\% & 72.22\% & 40\% \\ 
\textbf{Min} & -36\% & 0\% & -38.89\% & 0\% \\ 
\textbf{Standard Deviation} & 29.22\% & 18.94\% & 42.19\% & 15.06\% \\
\bottomrule
\end{tabular}
\end{table}

One expert think it is difficult to consider some responses in P5 as useful ideas. For example, responses in the following might be too unrealistic for designing a flying car.

\begin{itemize}
\item \textit {Augmented Reality (AR) Telepresence: using AR and virtual presence technology to reduce the need for physical travel, allowing teachers to conduct classes remotely while providing an immersive experience.}
\item \textit {Teleportation Technology (Theoretical): development of quantum teleportation for instant travel, though currently highly theoretical.} 
\item \textit {Suborbital Space Flights: using suborbital spacecraft for very fast point-to-point travel reducing intercontinental travel times to minutes.}
\end{itemize}

\begin{table}[H]
\caption{Selected high original (above 8) responses from P1 to P7 }
\begin{tabular}{lll}
\toprule
\textbf{NO} & \textbf{Responses} & \textbf{ Prompts with Similar Responses} \\
\midrule
{\textbf{\textit{P3}}} & \parbox[t]{9cm}{Balloon Lift: Big helium balloons could lift the car up, and small fans could move it around. The balloons could deflate when the car needs to land.} & ~ \\ \hline

\textbf{~} & \parbox[t]{9cm}{Rocket Boosters: Small rockets could be attached to the car for quick take-offs and landings. } & \textit{P4} \\  \hline

\textbf{\textit{P4}} & \parbox[t]{9cm}{Inflatable wings: The car could have wings that inflate when needed and deflate when it’s on the ground.} & \textit{P7} \\ \hline

\textbf{~} & \parbox[b]{10cm}{Magic carpets: Floating carpets that fly teachers to school (if magic was real!).} & ~ \\  \hline

\textbf{\textit{P7}} & \parbox[t]{9cm}{Magnetic Levitation (Maglev) Vehicles\\Portable Maglev Pads: Deployable maglev pads at schools and key locations for take-off and landing.} & \textit{P2}, \textit{P3} \\  \hline

\textbf{~} & \parbox{9cm}{Hybrid Airships\\Helium-Filled Airships: Using helium for lift, combined with small propulsion units for maneuvering, offering a safe and efficient alternative.} & ~ \\

\bottomrule
\end{tabular}
\end{table}

Table 15 shows a few responses from P1 to P7, all of which received high originality ratings. Some responses, such as inflatable wings, seem even impractical but they can inspire designers to think out of the box. We particularly noticed that the strategy – unconventional role can provide this kind of unexpected creative response. Despite different strategies being used, similar results appear, such as “rocket boosters” mentioned in both P3 and P4. The ideas do not receive high original rating due to the various reasons. Firstly, the ideas detected in both response and prompt, like P2’s one-shot example, are not considered original because ChatGPT can expand on the given prompt. Second, ChatGPT’s suggestions like folding wings, solar power, hydrogen fuel cell battery, magnetic track, or drone can be envisioned in the real-world and therefore were rated with lower originality values. We also notice that many responses mention about the infra-structure, regulations, or software systems. While these responses may not be high original, they can assist designer in thinking more thoroughly during the brainstorming, thereby enriching the design or expanding their idea range.

\section{Conclusion}
The advancement of large language models (LLMs) has significantly promoted the utilization of prompts in various domains. However, existing methods of creating prompts focus solely on increasing response accuracy. There has been a notable gap in addressing creativity in prompt design and evaluating its output. To address this gap, we developed a GPS framework for cooperating with an LLM tool in brainstorming. The framework is designed to augment creativity. \\

Our study demonstrated the application of the GPS framework. Through our evaluation, we discovered that using the GPS template resulted in increased values across all dimensions. Implementing these strategies aids designers in enhancing flexibility, originality, and elaboration. Consequently, our GPS framework proves effective in boosting creativity when designers collaborate with ChatGPT 3.5. We also identified that the GPS framework not only supports the generation of more diverse and original ideas but also enhances the elaboration of responses. Our GPS framework ensures that creativity is systematically fostered, offering a tool for designers seeking to increase the creative responses from their prompts.












\bibliographystyle{unsrtnat}
\bibliography{references}  
\nocite{*}






\end{document}